\documentclass[aps,prb,twocolumn,superscriptaddress]{revtex4}
\usepackage{graphicx}
\usepackage{graphics}
\usepackage{color}
\usepackage{ulem}

\begin{document}

\title{Reciprocity in diffusive spin-current circuits}

\author{Ya.\ B. Bazaliy}
  \affiliation{University of South Carolina, Columbia SC 29208, USA}
  \email{yar@physics.sc.edu}
\author{R. R. Ramazashvili}
 \affiliation{Laboratoire de Physique Th\'eorique, Universit\'e de Toulouse, CNRS, UPS, France}
 \email{revaz@irsamc.ups-tlse.fr}
\date{\today}

\begin{abstract}
Similarly to their purely electric counterparts, spintronic circuits may be presented as networks of lumped elements. Due to interplay between spin and charge currents, each  element is described by a matrix conductance. We establish reciprocity relations  between the entries of the conductance matrix of a multi-terminal linear device, comprising normal metallic and strong ferromagnetic elements with spin-inactive interfaces between them. In particular, reciprocity equates the spin transmissions through a two-terminal element in the opposite directions. When applied to ``geometric spin ratchets'', reciprocity shows that certain effects, announced for such devices, are, in fact, impossible. Finally, we discuss the relation between our work and the spintronic circuit theory formalism.
\end{abstract}

\maketitle

\section{Introduction}

Spin currents have been actively discussed in the context of spintronics, a field where memory and logic devices employ electron spin on a par with its charge. A number of theoretical concepts have been developed to describe operation of such devices. As the field matures, one needs to build and work with ever larger networks of connected spintronic elements -- akin to how electric circuits are composed of elementary resistors, capacitors {\it etc.} To this end, a spin circuit theory was proposed in the pioneering paper Ref.~\onlinecite{nazarov:2000}. The principles of the latter approach were then used to formulate circuit descriptions, that may be more convenient for applications. \cite{manipatruni:2012, camsari:2015}

In DC electric circuits a textbook resistor is characterized by a single parameter, the resistance $R$ that encapsulates the element's material properties, shape, size, and contact positions. In spintronics, where spin and charge currents are interconnected, even the simplest element is characterized not by a single number but by a conductance matrix.\cite{manipatruni:2012, camsari:2015}

In this paper we show -- within the assumptions detailed below -- that the entries of a spintronic conductance matrix obey certain general relations that are  independent of shape, size and material constants of the actual physical elements, and are similar to classic reciprocity relations for electric circuits.\cite{carson:1924, panofsky-phillips, wikipedia_on_reciprocity}

The ultimate goal of a circuit theory is to describe spintronic circuits using generalized Kirchhoff's rules. Realizing this program, one has to keep in mind, however, that certain differences between spin and electric currents invalidate much of the intuition accumulated in electric circuits. First, unlike electric current, spin current is not conserved, and in a two-terminal element the incoming and outgoing spin currents are generally different: An element cannot be characterized by a single value of spin current. Second, spin currents behave differently from electric currents when potentials applied to the two terminals of an element are interchanged. For electric current, the Ohm's law $I = G (V_1 - V_2)$, expressed through the conductivity $G = 1/R$, states that, by interchanging $V_1$ and $V_2$, one flips the sign of the current, but preserves its magnitude. In this sense, resistor is a directionless element. As detailed below, this does not apply to two-terminal spin elements, where interchanging the terminal potentials generally changes the magnitudes of both incoming and outgoing spin currents. However, the relations between the entries of the conductance matrix, obtained in this paper, show that a two-terminal spin element behaves in a familiar way with respect to interchange of potentials in a special case, where a driving spin potential is applied to one terminal, and the resulting spin current is measured at the other, grounded, terminal. This means that the transmission of spin current through a spin-dissipating element is directionless.

\begin{figure}[b]
\center
\includegraphics[width = 0.48\textwidth]{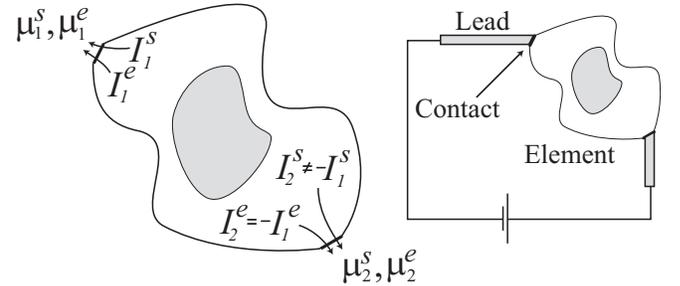}
\caption{Left: a two-terminal element, an island with two contacts, where spin potentials $\mu^s_{1,2}$ and electric potentials $\mu^e_{1,2}$ are applied. Conserved electric current ($I^e_1 = - I^2_2$) and non-conserved spin current ($I^s_1 \neq - I^s_2$) are shown schematically. White denotes a normal metal, gray denotes a ferromagnet. Right: the same connected to ferromagnetic leads (gray) in a spin circuit.}
 \label{fig:device}
\end{figure}

\section{Reciprocity in the diffusion regime}
We consider metallic devices in the diffusion regime, with the mean free paths of charge carriers being much shorter than any other length scale in the problem. In this approximation, the electron state is completely described by the distributions of electric potential $\mu^e(r)$ and spin potential ${\mu}^{s}(r)$.\cite{campbell:1967, valet-fert_prb1993, rashba_prbrc2000, rashba_epjb2002} Instead of electric current density $j_i$, we will work with particle current density $j^e_{i} = j_i/e$. Here index $i = \{x,y,z\}$ denotes direction in real space. Spin current density $j_{i}^{s\alpha}$ has two indices with $\alpha = \{x,y,z\}$ denoting direction in spin space. Spin current is also defined in terms of the number of particles: passage of one spin-up electron per second through a mathematical plane contributes one, not $1/2$, to $j^s$ flowing through it. In the diffusion regime, currents are driven by the gradients of electric potential $\mu^e(r)$ and spin potential $\mu^{s\alpha}(r)$.\cite{campbell:1967, valet-fert_prb1993, rashba_prbrc2000, rashba_epjb2002}

A sample device is shown in Fig.~\ref{fig:device}. The element has arbitrary shape and may contain magnetic and non-magnetic metal parts. Two contacts connect it to the outside world. They are assumed to be small enough for the electric and spin potentials to be considered constant across each of them.  In order to apply spin potentials to the element, the external circuit must involve magnetic elements, producing the required spin imbalance.

\subsection{A two-terminal normal metal element}\label{sec:Normal_metal_elements}
We start with a conceptually simpler case of a normal metal element. The currents
are related to the potentials as per
\begin{eqnarray}
\label{eq:normal_metal_electric_current}
  j^e_i &=& -\frac{\sigma}{e^2} \nabla_i \mu^e \ ,
  \\
\label{eq:normal_metal_spin_current}
  j^{s\alpha}_i &=& -\frac{\sigma}{2 e^2} \nabla_i \mu^{s\alpha},
\end{eqnarray}
where $\sigma$ is the (possibly non-uniform) electric conductivity of the metal. We will study steady-state solutions, where the continuity equation
$\partial_t \rho + {\bf \nabla \cdot j} = 0$ yields
\begin{equation}\label{eq:electric_current_conservation}
\nabla_i j^e_i = 0
\end{equation}
for the electric current, and
\begin{equation}\label{eq:normal_metal_spin_relaxation}
\nabla_i j^{s \alpha}_i = \frac{\nu}{\tau_s}\mu^{s \alpha} ,
\end{equation}
for the spin current, with $\nu$ being the density of states of the normal metal, and $\tau_s$ the spin relaxation time.

Equations (\ref{eq:normal_metal_electric_current},\ref{eq:electric_current_conservation}) for electric potential, and (\ref{eq:normal_metal_spin_current},\ref{eq:normal_metal_spin_relaxation}) for spin potential are decoupled. Once the system (\ref{eq:normal_metal_spin_current},\ref{eq:normal_metal_spin_relaxation}) is solved,
the spin current density $j^{s\alpha}_i$ can be found everywhere, and the total spin current flowing thorough each contact is given by
$$
I^{s\alpha}_{t} = \int_{S_t} j^{s\alpha}_i dA_i ,
$$
where the integration goes over the contact surface $S_t$, and $t = 1,2$ labels the two contacts. It is, or course, assumed that the spin current does not leak in or out anywhere else at the sample boundary. By definition, the current is considered positive if it flows out of the element, i.e., surface element $d{\bf A}$ points along the outward
normal. Due to the linearity of the Eq.~(\ref{eq:normal_metal_spin_current}), the
total spin currents must be linearly related to the spin potentials of the terminals
\begin{equation}\label{eq:normal_metal_spin_conductance}
I^{s\alpha}_{t} = G^s_{tt'} \mu^{s\alpha}_{t'}
\end{equation}
via the matrix spin conductance $G^s_{tt'}$, that is determined by the solution of
the system (\ref{eq:normal_metal_spin_current}),(\ref{eq:normal_metal_spin_relaxation}). Since both equations in the system are diagonal in the spin index $\alpha$, the conductance is diagonal in it as well. We will thus suppress the spin index in the equations for normal metal elements.

Note that a purely electric two-terminal device can be described by a matrix conductance similarly to the Eq.~(\ref{eq:normal_metal_spin_conductance}). However, the Ohm's law $I_{2} = - I_{1} = G(V_1 - V_2)$ constrains the electric conductance matrix to a form
$$
\hat G^e = \left| \begin{array}{rr} -G & G \\  G & -G \end{array} \right| ,
$$
with a single independent entry. The spin conductance matrix
$$
\hat G^s = \left| \begin{array}{cc} G^s_{11} & G^s_{12} \\  G^s_{21} & G^s_{22} \end{array} \right|
$$
has four entries, and one may ask whether there are any relations between them that hold regardless of the shape and material of the spintronic element.

We now prove that the answer to the question above is affirmative, and the off-diagonal ele\-ments of $\hat G^s$ are always equal. The proof is based on the so-called reciprocity property \cite{panofsky-phillips} of the solutions of the Eq.~(\ref{eq:normal_metal_spin_relaxation}), summarized in Appendix~\ref{appendix:identities}. Imagine solving this equation for mixed boundary conditions, specified by constant spin potentials $\mu^s(r) = \mu^s_t$ at the contacts, and $j^s_i n_i = 0$ outside the contact areas (no current penetrating the boundary) where $n_i$ is the local normal to the surface of the element. Consider two solutions, each for a se\-pa\-ra\-te pair of potentials $\mu^s_t$ applied at the contacts $t = 1,2$. These solutions will be denoted $\mu^s(r,c)$, with a ``case label'' $c = 1,2$. Knowing the $\mu^s(r,c)$, one can find the currents $j^{s}_i(r,c) = - (\sigma/2e^2) \nabla_i \, \mu^{s}(r,c)$. Now, let us use the functions $\mu^s(r,c)$ and $j^{s}_i(r,c)$ to calculate the integral
$$
Q  =  \int \big[ \mu^s(r,1) \nabla_i j^s_i(r,2) - \mu^s(r,2) \nabla_i j^s_i(r,1) \big] dV \ .
$$
On the one hand, the Eq.~(\ref{eq:normal_metal_spin_relaxation}) tells us that $Q = 0$.
On the other hand, the identity (\ref{eq:identity_scalar}) of Appendix~\ref{appendix:identities} transforms $Q$ into the surface integral
\begin{eqnarray*}
Q &=& \oint \big[
\mu^s(r,1) \big( - \sigma \nabla_i \mu^s(r,2) \big)
\\
&-& \mu^s(r,2) \big( - \sigma \nabla_i \mu^s(r,1) \big)
\big] dA_i .
\end{eqnarray*}
Since spin potentials are constant across each contact, the contacts do not overlap,
and $j^s_i$ crosses the surface only at the contacts, we obtain
$$
Q = \mu^s_t(1) I^s_t(2) - \mu^s_t(2) I^s_t(1) = 0
$$
with summation over repeating indices $t$. Expressing currents through
potentials via (\ref{eq:normal_metal_spin_conductance}), we find
$$
\mu^s_t(1) G^s_{tt'} \mu^s_{t'}(2) - \mu^s_t(2) G^s_{tt'} \mu^s_{t'}(1) = 0
$$
or
$$
\mu^s_t(1) \mu^s_{t'}(2) (G^s_{tt'} - G^s_{t't}) = 0.
$$
As the potentials $\mu^s_t(1)$ and $\mu^s_t(2)$ can be chosen arbitrarily,
the above means $G^s_{tt'} = G^s_{t't}$. Thus, a $2 \times 2$ matrix $G^s_{tt'}$
obeys the constraint
\begin{equation}
\label{eq:normal_element_main_result}
G^s_{12} = G^s_{21}.
\end{equation}

The physical meaning of the Eq.~(\ref{eq:normal_element_main_result}) manifests itself in experiments with a single driving terminal, i.e. $\mu^s_t(c) = \mu^{s0} \delta_{tc}$. In the first case ($c = 1$), spin potential $\mu^{s0} > 0$ is applied to the first (driving) terminal, while the second terminal is kept at zero spin potential (ground terminal). In the second case ($c = 2$), the driving terminal and the ground terminal are interchanged. It is physically clear (and can be mathematically proven) that, in the first case, a current $(-I^s_{1}) > 0$ will enter the element at the driving terminal one, and a current $I^s_{2}>0$ will leave it at the ground terminal. As already discussed, the transmitted current will be smaller due to spin dissipation: $(-I^s_{1}) > I^s_{2}$. Equation~(\ref{eq:normal_metal_spin_conductance}) now reads $I^{s}_{2} = G^s_{21} \mu^{s0}$, i.e., the matrix conductance element $G^s_{21}$ parameterizes the transmission of spin current through the device from the driving terminal to the ground terminal. In the second case, spin current is driven by the same spin potential, applied to terminal two. The current, transmitted from the driving terminal to the ground terminal is now given by $I^{s}_{1} = G^s_{12} \mu^{s0}$. The reciprocity property $G^s_{12} = G^s_{21}$ means that, for equal potentials applied to the driving terminal, the spin current transmitted to the ground terminal is independent of which of the two terminals is driven. In other words, spin transmission from the driving terminal to the ground terminal is directionless.

\begin{figure}[t]
\center
\includegraphics[width = 0.25\textwidth]{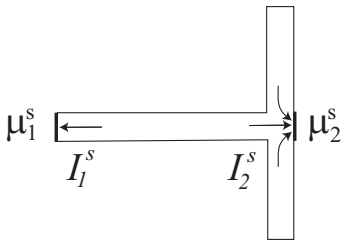}
\caption{Asymmetric normal element.}
 \label{fig:device_asymmetric}
\end{figure}

We now show that the diagonal elements of $\hat G^s$ may differ from each other. Consider a geometrically asymmetric element such as the one in Fig.~\ref{fig:device_asymmetric}. Apply potentials $\mu^s_t(c) = \mu^{s0} \delta_{tc}$ as discussed above, and measure the current flowing through the driving terminal. On the one hand, it equals $I^{s}_{1}(1) = \mu^{s0} G^s_{11}$ in the first case, and $I^{s}_{2}(2) = \mu^{s0} G^s_{22}$ in the second. On the other hand, it is physically clear that for such an element these currents are different, since spins injected into the $t=1$ terminal can diffuse in only one direction, whereas spins injected into the $t=2$ terminal can also diffuse into the vertical bar (cf. Ref.~\onlinecite{kimura_prb2005}, Sec.~IIIC), thus increasing the total spin current entering the element. Due to our definition of current signs, we have $I^{s}_{2}(2) < I^{s}_{1}(1) < 0$. As a result, $G^s_{22} < G^s_{11} < 0$. Put more generally, in an asymmetric element of a size exceeding the spin diffusion length, the diagonal elements $G^s_{11}$ and $G^s_{22}$ are primarily defined by the geometry and material properties of the device within a few diffusion lengths from the corresponding contact.

The reciprocity Eq.~(\ref{eq:normal_element_main_result}) presents the main
result of our work in the simplest setting of a two-terminal diffusive normal-metal element: Transmission of spin current between the terminals is direction-independent. We now proceed to describe the reciprocity relations that emerge in more
general settings.

\subsection{Composite elements, incorporating normal metals and strong ferromagnets}
\label{sec:Composite_elements}
\begin{figure}[t]
\center
\includegraphics[width = 0.35\textwidth]{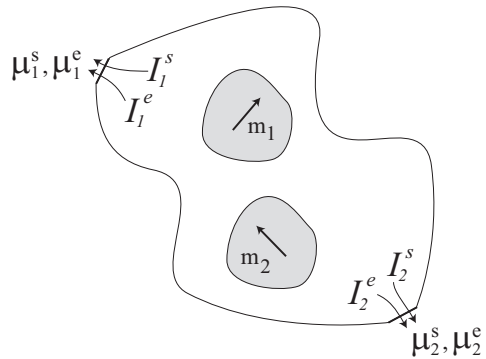}
\caption{Two-terminal element with normal (white) and ferromagnetic (shaded) parts.}
 \label{fig:element_FM}
\end{figure}

In this section, we will consider a two-terminal element, comprising normal ferromagnetic ($F$)
as well as non-magnetic ($N$) regions (Fig.~\ref{fig:element_FM}). In each ferromagnetic region, magnetization is assumed to be uniform; magnetizations of different $F$ regions are not expected to be collinear. We restrict our analysis to strong ferromagnets, where itinerant electron spins are polarized along the direction of local magnetization. It is further assumed that different ferromagnetic parts do not border each other directly, but are always separated by a non-magnetic region. The boundaries between the normal and ferromagnetic regions are assumed to be Ohmic (no tunnel barriers).

Inside a strong ferromagnet, the spin potential may be presented as $\mu^{s\alpha}(r) = \mu^s(r) m^{\alpha}$, where $m^{\alpha}$ is a unit vector along the magnetization. The currents are given by expressions \cite{campbell:1967, valet-fert_prb1993, rashba_prbrc2000, rashba_epjb2002}
\begin{eqnarray}
\label{eq:strong_ferromagnet_electric_current}
j^e_i &=& - \frac{\sigma}{e^2} \left(  \nabla_i\mu^e + \frac{1}{2}p\nabla_i\mu^s  \right),
\\
\label{eq:strong_ferromagnet_spin_current}
j^{s\alpha}_i &=& -   \frac{m^{\alpha} \sigma}{e^2} \left(  \frac{1}{2}\nabla_i\mu^s +  p \nabla_i\mu^e \right),
\end{eqnarray}
where $p$ is the spin polarization parameter, characterizing the material of a given ferromagnetic part. Since $j^{s\alpha}_i \propto m^\alpha$, it follows that $(\delta^{\alpha\beta} - m^{\alpha} m^{\beta}) j^{s\beta} = 0$.

To streamline the formulas, we combine electric and spin potentials into a four-component rescaled potential $\tilde\mu^a(r) = \{\mu^e, \mu^{sx}/2, \mu^{sy}/2, \mu^{sz}/2 \}$, where $a = \{e, sx, sy, sz\}$. Likewise, the currents are combined into $j_i^a = \{j^e_i, j^{sx}_i, j^{sy}_i, j^{sz}_i \}$. Then, the Eqs.~(\ref{eq:strong_ferromagnet_electric_current}) and (\ref{eq:strong_ferromagnet_spin_current}) take the form
\begin{equation}\label{eq:notations}
j_i^a = -\Sigma^{ab} \nabla_i \tilde\mu^{b} ,
\end{equation}
where $\Sigma$ is the generalized conductivity matrix. The use of $\tilde\mu^a$ renders $\Sigma^{ab}$ symmetric (Appendix~\ref{Appendix:Conductivty_of_strong_ferromagnets}), which allows us to apply the identity (\ref{eq:identity_tensor})
\begin{eqnarray}
\nonumber
Q & \equiv & \int \big( \tilde\mu^a(1) \nabla_i j^a_i(2) - \tilde\mu^a(2) \nabla_i j^a_i(1)\big) dV
\\
\label{eq:identity_unified_electric_spin_current_potentials}
&=& \oint \big( \tilde\mu^a(1) j^a_i(2) - \tilde\mu^a(2) j^a_i(1)\big) dA_i \ .
\end{eqnarray}
The divergences $\nabla_i j^a_i$ in the volume integral are non-zero for the spin part only. In the bulk, be it normal or ferromagnetic, one has $\nabla_i j^a_i = \{ 0, \nu(r) \mu^{\alpha}/\tau(r) \}$ with material-specific effective densities of states and relaxation times.\cite{rashba_prbrc2000, rashba_epjb2002} A direct check shows that bulk relaxation gives zero contribution to $Q$.

However, in a composite device spin relaxation is not limited to the bulk  but acquires an  additional contribution from the $F/N$ interfaces. Here we will assume Ohmic, spin-inactive interfaces: At the interface $S$, the potentials are continuous
\begin{equation}\label{eq:bc_on_mu}
\left. \mu^a(N) \right|_S = \left. \mu^a(F) \right|_S = \{ \mu^e, \mu^s m_x, \mu^s m_y, \mu^s m_z \}
\end{equation}
but the currents $j^a$ are not.\cite{brataas:2001, brataas:2006} Spin current may have arbitrary direction in spin space on the normal-metal side of the interface, but has to be parallel to $m^{\alpha}$ on its ferromagnetic side. The spin current component perpendicular to $m^{\alpha}$ is absorbed in a thin boundary layer near the interface, while the current component parallel to $m^{\alpha}$ is continuous. In the strong-ferromagnet approximation the absorption layer thickness is infinitesimally small, so, the boundary conditions for currents read
\begin{eqnarray}
\label{eq:electric_boundary_condition}
&& j^e_i(N) n_i \big|_S = j^e_i(F) n_i \big|_S,
 \\
\label{eq:spin_boundary_condition} &&  m^{\alpha} j^{s\alpha}_i(N) n_i \big|_S = m^{\alpha} j^{s\alpha}_i(F) n_i  \big|_S ,
\end{eqnarray}
with $n_i$ being the normal to the interface. The discontinuity of the perpendicular spin current gives rise to a surface absorption term
\begin{equation}
\label{eq:surface_absorption}
\nabla_i j^{s\alpha}_i = R^{\alpha}(r) =   \big( \delta^{\alpha\beta} - m^{\alpha} m^{\beta} \big) j^{s\beta}_i(N) \, n_i \delta_{S}(r)
\end{equation}
proportional to the surface delta-function $\delta_{S}$ at the $F/N$ interface. In the expression (\ref{eq:surface_absorption}) the spin current is evaluated on the normal metal side of the interface.

We now show that $Q$ also vanishes in the presence of surface absorption (\ref{eq:surface_absorption}). Indeed, since spin potential is continuous at the N/F interface, $\mu^{s\alpha} = m^{\alpha} \mu^s$ on both sides of the surface, and
$$
\tilde\mu^{s\alpha}(c) \nabla_i j^{s\alpha}_i(c') =
  \frac{1}{2} m^{\alpha}\mu^s(c) R^{\alpha} = 0
$$
where we used $m^{\alpha} ( \delta^{\alpha\beta} - m^{\alpha} m^{\beta}) = 0$. We therefore have
$$
\oint \big( \tilde\mu^a(1) j^a_i(2) - \tilde\mu^a(2) j^a_i(1)\big) dA_i = 0,
$$
and hence after integration
$$
\tilde\mu^a_t(1) I^a_t(2) - \tilde\mu^a_t(2) I^a_t(1) = 0 .
$$
In a composite two-terminal element the Eq. (\ref{eq:normal_metal_spin_conductance}) is generalized to
\begin{equation}\label{eq:composite_element_conductance}
I^a_t = G^{ab}_{tt'} \tilde\mu^b,
\end{equation}
and thus
$$
\tilde\mu^a_t(1) \, {G}^{ab}_{tt'} \, \tilde\mu^b_{t'}(2) -
\tilde\mu^a_t(2) \, {G}^{ab}_{tt'} \, \tilde\mu^b_{t'}(1) = 0,
$$
or
$$
\tilde\mu^a_t(1) \tilde\mu^b_{t'}(2) \big(
{G}^{ab}_{tt'} - {G}^{ba}_{t't}
\big) = 0.
$$
Since we are free to choose the potentials $\tilde\mu^a_t(1)$ and $\tilde\mu^a_t(2)$ arbitrarily, the above equality means
\begin{equation}\label{eq:composite_element_conductance_condition}
{G}^{ab}_{tt'} = {G}^{ba}_{t't}.
\end{equation}
This equation generalizes our result (\ref{eq:normal_element_main_result})
to a composite two-terminal diffusive element.

The symmetry requirement (\ref{eq:composite_element_conductance_condition}) applied to an $n \times n$ matrix produces $n(n-1)/2$ relations between its entries. For an $8 \times 8$ matrix ${G}^{ab}_{tt'}$ this yields 28 relations between 64 entries. Note that relations between the elements with $t \neq t'$ and $a=b$ have a meaning similar to that of (\ref{eq:normal_element_main_result}): Spin transmission from one contact to another is direction-independent. In particular, for $a = b = e$ one recovers the direction-independence of the charge transport, already well-known from elementary physics. For $a=e$ and $b = x,y,z$ we find a relation between the spin currents generated by the electric potential, and vice versa.

\subsection{Multi-terminal elements}
An element with $N$ terminals is described by the conductance $G^{ab}_{tt'}$ with $t, t' = 1 \ldots N$: Conductance is a $4N \times 4N$ matrix. Applying the procedure of the previous section, we can prove the relation
$$
\mu^a_t(1)\mu^b_{t'}(2) \big( G^{ab}_{tt'} - G^{ba}_{t't} \big) = 0
$$
for any choice of $4N$-dimensional vectors $\mu^a_t(1)$ and $\mu^a_t(2)$. Thus the Eq.~(\ref{eq:composite_element_conductance_condition}) holds for $G^{ab}_{tt'}$, and the $4N \times 4N$ conductance matrix $G^{ab}_{tt'}$ is symmetric as well.

\section{Consequences of electric current conservation}

\subsection{Two-terminal elements}
Let us return to the two-terminal case. The tensor ${G}^{ab}_{tt'}$ may be represented as an $8 \times 8$ matrix in two ways. Firstly as
\begin{equation}\label{eq:G4x4blocks}
\left( \begin{array}{c} I^a_1 \\ I^a_2 \end{array}\right) =
\left|\begin{array}{c|c}
G^{ab}_{11} & G^{ab}_{12}
\\ \hline
G^{ab}_{21} & G^{ab}_{22}
\end{array}\right|
\left( \begin{array}{c} \tilde\mu^b_1 \\ \tilde\mu^b_2 \end{array}\right) ,
\end{equation}
with $4 \times 4$ matrix entries in every block, secondly as
\begin{equation}\label{eq:G2x2blocks}
\left( \begin{array}{c} I^e_t \\ I^{sx}_t \\ I^{sy}_t \\ I^{sz}_t \end{array}\right) =
\left|\begin{array}{c|c|c|c}
G^{ee}_{tt'} & G^{e,sx}_{tt'} & G^{e,sy}_{tt'} & G^{e,sz}_{tt'}  \\
\hline
G^{sx,e}_{tt'} & G^{sx,sx}_{tt'} & G^{sx,sy}_{tt'} & G^{sx,sz}_{tt'}  \\
\hline
G^{sy,e}_{tt'} & G^{sy,sx}_{tt'} & G^{sy,sy}_{tt'} & G^{sy,sz}_{tt'}  \\
\hline
G^{sz,e}_{tt'} & G^{sz,sx}_{tt'} & G^{sz,sy}_{tt'} & G^{sz,sz}_{tt'}
\end{array}\right|
\left( \begin{array}{c} \mu^{e}_{t'} \\ \tilde\mu^{sx}_{t'} \\ \tilde\mu^{sy}_{t'} \\ \tilde\mu^{sz}_{t'} \end{array}\right) ,
\end{equation}
with $2 \times 2$ matrices in every block. The order of the tensor's elements in two cases is different but in both representations the resulting $8 \times 8$ matrix is symmetric.

The second representation is more convenient for taking into account the electric current conservation: For any set of applied potentials $I^e_1 = -I^e_2$ (the minus on the left hand side appears due to our definition of current signs). This gives $G^{ea}_{1t} = - G^{ea}_{2t}$ for every $t$ and $a$, where, in the notation of the Eqs. (\ref{eq:notations}), (\ref{eq:G4x4blocks}) and (\ref{eq:G2x2blocks}), $a$ takes the values $e$, $sx$, $sy$ and $sz$. This amounts to eight more constraints on the entries of the conductance matrix, which further reduces the number of its independent entries to 28 = 64 - 28 (reciprocity) - 8 (electric current conservation).

Note that, together with the reciprocity condition, the electric current conservation yields $G^{ae}_{t1} = - G^{ae}_{t2}$. That is, every current $I^a_t$ depends only on the difference $\mu^e_1 - \mu^e_2$, as required by gauge invariance. This is simply a manifestation of the intimate relation between gauge invariance and charge conservation.\cite{schrieffer1988} Equi\-va\-lent\-ly, we could impose gauge invariance via $G^{ae}_{t1} = - G^{ae}_{t2}$, which would then imply eight constraints and, of course, yield electric current conservation. Needless to say, the conductance matrix ends up with the same 28 independent entries.

With current conservation taken into account,
$$
G^{ee}_{tt'} =
\left|\begin{array}{rr}
-G & G
\\
G & -G
\end{array}\right|, \quad
G^{s\alpha,e}_{tt'} =
\left|\begin{array}{rr}
C^{\alpha}_1 & C^{\alpha}_2
\\
-C^{\alpha}_1 & -C^{\alpha}_2
\end{array}\right|.
$$

Now we can write down the $8 \times 8$ matrix $G^{ab}_{tt'}$ that obeys all the constraints and has 28 independent entries
\begin{widetext}
\begin{equation}\label{eq:8x8_explicit_matrix}
\left( \begin{array}{c} I^e_1 \\ I^e_2 \\ I^{sx}_1 \\ I^{sx}_2 \\ I^{sy}_1 \\ I^{sy}_2 \\
I^{sz}_1 \\ I^{sz}_2 \end{array}\right) =
\left|\begin{array}{rrrrrrrr}
- G & G & C^{x}_1 & C^{x}_2 & C^{y}_1 & C^{y}_2 & C^{z}_1 & C^{z}_2
\\
G & -G & -C^x_1 & -C^{x}_2 & -C^{y}_1 & -C^{y}_2 & -C^{z}_1 & -C^{z}_2
\\
C^{x}_1 & -C^{x}_1 & S^{x}_1 & S^{x}_c & S^{xy}_{11} & S^{xy}_{12} & S^{xz}_{11} & S^{xz}_{12}
\\
C^{x}_2 & -C^{x}_2 & S^{x}_c & S^{x}_2 & S^{xy}_{21} & S^{xy}_{22} & S^{xz}_{21} & S^{xz}_{22}
\\
C^{y}_1 & -C^{y}_1 & S^{xy}_{11} &  S^{xy}_{21} & S^y_1 & S^y_c & S^{yz}_{11} & S^{yz}_{12}
\\
C^{y}_2 & C^{y}_2 & S^{xy}_{12} & S^{xy}_{22} & S^y_c & S^y_2 & S^{yz}_{21} & S^{yz}_{22}
\\
C^{z}_1 & -C^{z}_1 & S^{xz}_{11} & S^{xz}_{21} & S^{yz}_{11} & S^{yz}_{21} & S^z_1 & S^z_c
\\
C^{z}_2 & -C^{z}_2 & S^{xz}_{12} & S^{xz}_{22} & S^{yz}_{12} & S^{yz}_{22} & S^z_c & S^z_2
\end{array}\right|
\left( \begin{array}{c} \mu^e_1 \\ \mu^e_2 \\ \tilde\mu^{sx}_1 \\ \tilde\mu^{sx}_2 \\ \tilde\mu^{sy}_1 \\ \tilde\mu^{sy}_2 \\
\tilde\mu^{sz}_1 \\ \tilde\mu^{sz}_2 \end{array}\right)
\end{equation}
\end{widetext}
For electric current we obtain the expression
\begin{equation}\label{eq:composite_element_electric_current}
I^e_t =  (-1)^t \big[ G(\mu^e_1 - \mu^e_2) + C^{\alpha}_{t'} \tilde\mu^{s\alpha}_{t'} \big] .
\end{equation}
For spin current we find
\begin{equation}\label{eq:composite_element_spin_current}
I^{s\alpha}_t = C^{\alpha}_t (\mu^e_1 - \mu^e_2) +  S^{\alpha\beta}_{tt'} \tilde\mu^{s\beta}_{t'}
\end{equation}
with a symmetric spin conductance matrix $S^{\alpha\beta}_{tt'} = S^{\beta\alpha}_{t't}$. The matrix elements of $S^{\alpha\beta}_{tt'}$ with $\alpha \neq \beta$, $t \neq t'$ describe the precession of spin injected at one terminal while it is transmitted to the other terminal. For instance, $S^{xy}_{12}$ describes spin precession from $y$ to $x$ that may occur due to, e.g., the presence of magnetic parts in the element. Note that the reciprocity relations do not connect the $S^{xy}_{12}$ and $S^{xy}_{21}$, that is transmission in the opposite spatial directions with the same spin precession. Instead, the equation $S^{xy}_{12} = S^{yx}_{21}$ connects the processes that are opposite in both the spatial direction and the sense of spin precession.

\begin{figure}[b]
  \centering
  \includegraphics[width = 0.3\textwidth]{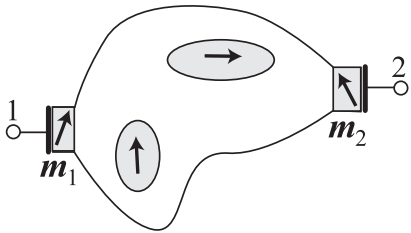}
  \caption{Composite element with ferromagnetic contacts.}
  \label{fig:element_F-terminals}
\end{figure}

An interesting special case is found when both terminals of an element are strong ferromagnets (Fig.~\ref{fig:element_F-terminals}). The magnetization directions of the terminals, $m^{\alpha}_1$ and $m^{\alpha}_2$, may be non-collinear. Spin potentials and spin currents at the terminals are restricted to the form $\mu^{s\alpha}_{1} = m^{\alpha}_{1} \mu^s_{1}$ and $\mu^{s\alpha}_{2} = m^{\alpha}_{2} \mu^s_{2}$, $I^{s\alpha}_1  = m^{\alpha}_{1} I^s_1$ and $I^{s\alpha}_2  = m^{\alpha}_{2} I^s_2$. The Eq.~(\ref{eq:8x8_explicit_matrix}) then reduces to a simpler one, involving a $4 \times 4$ conductance matrix as per
\begin{equation}\label{eq:4x4_matrix_for_F_terminals}
\left( \begin{array}{c} I^e_1 \\ I^e_2 \\ I^{s}_1 \\ I^{s}_2  \end{array}\right) =
\left|\begin{array}{rrrr}
- G & +G & C_1 & C_2
\\
G & -G & -C_1 & -C_2
\\
C_1 & -C_1 & S_1 & S_c
\\
C_2 & -C_2 & S_c & S_2
\end{array}\right|
\left( \begin{array}{c} \mu^e_1 \\ \mu^e_2 \\ \tilde\mu^{s}_1 \\ \tilde\mu^{s}_2  \end{array}\right)
\end{equation}
with $C_1 = C^{\alpha}_1 m^{\alpha}_1$, $C_2 = C^{\alpha}_2 m^{\alpha}_2$, $S_1 = m^{\alpha}_1 S^{\alpha\beta}_{11} m^{\beta}_1$, $S_2 = m^{\alpha}_2 S^{\alpha\beta}_{22} m^{\beta}_2$, and $S_c = m^{\alpha}_1 S^{\alpha\beta}_{12} m^{\beta}_2$. We see that the conductance of such an element is defined by 6 independent parameters.

A related special case, admitting an equally simple description, is found when the magnetizations of all ferromagnetic parts of a composite element are collinear.
The terminals may be ferromagnetic or normal, but it is required that the
applied spin potentials are collinear with the magnetization direction.
The situation then reduces to the one described by
Eq.~(\ref{eq:4x4_matrix_for_F_terminals}) with $m^{\alpha}_1 = m^{\alpha}_2$.

\begin{figure}[t]
  \centering
  \includegraphics[width = 0.4\textwidth]{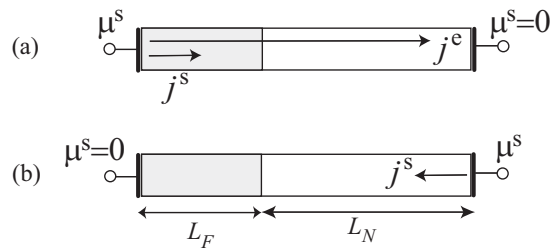}
  \caption{Composite element consisting of a ferromagnetic (shaded) and normal metal (white) parts. Electric current is generated by spin potential applied to one terminal (a) but not the other (b).}
  \label{fig:element_Non-equal_C1_C2}
\end{figure}

Returning to the general expressions (\ref{eq:composite_element_electric_current}) and (\ref{eq:composite_element_spin_current}) for the currents we stress that, unlike the electric potentials $\mu^e_t$, spin potentials do not have to appear only in the form of a difference $\mu^{s\alpha}_1 - \mu^{s\alpha}_2$. In other words, the coefficients $C^{\alpha}_1$ and $C^{\alpha}_2$ are not necessarily equal in absolute value and opposite in sign. The same is true for $S^{\alpha \beta}_{t1}$ and $S^{\alpha \beta}_{t2}$. The absence of such a requirement becomes transparent in a collinear setup, where $\mu^{s\alpha}$ differs from zero for a single direction $\alpha$ in spin space. Here it is evident that $\mu^s_t = \mu^\uparrow_t - \mu^\downarrow_t$ is already
gauge-invariant for each $t$ as the difference of spin-up and spin-down potentials. Thus electric and spin currents may depend separately on $\mu^s_1$ and $\mu^s_2$ without violating gauge invariance.

We illustrate this point by an explicit example of a composite element, consisting of ferromagnetic and normal parts, shown in the Fig.~\ref{fig:element_Non-equal_C1_C2}. The parts have lengths $L_{F,N}$, much larger than the spin diffusion lengths $\lambda_{F,N}$ in either material. We assume that the ferromagnet is magnetized along the $x$ direction. No electric potentials are applied to the element.

In the first experiment, a spin potential is applied along the magnetization, to the left ($t=1$) contact only, $\mu^{s\alpha}_t = \mu^{s0} \delta^{\alpha x}\delta_{t1}$. Thus spin current $j^{sx}$ is injected and propagates along the ferromagnetic part of the element over a distance $\lambda_F$ before dissipating. The presence of non-zero $j^{sx}$ in a ferromagnet, in turn, generates electric current $j^e$ according to the Johnson-Silsbee physics.\cite{johnson-silsbee_prl1985, takahashi_prb2003} Electric current, once generated, reaches the right terminal of the element. The total electric current is given by $I^e_1 = - I^e_2 = C^x_1 \mu^{s0} \neq 0$. Thus $C^x_1 \neq 0$.

In the second experiment, spin potential is only applied to the right ($t=2$) contact
$\mu^{s\alpha}_t = \mu^{s0}\delta^{\alpha x}\delta_{t2}$, also injecting spin current. As in the previous case, the spin current completely dissipates before reaching the boundary between the parts of the element. However, in a normal metal, pure spin current generates no electric current, and thus $I^e_1 = - I^e_2 = C^x_2 \mu^{s0} = 0$. Therefore, $C^x_2 = 0 \neq C^x_1$.

\subsection{Multi-terminal elements}
An easy way to find the additional constraints arising from electric current conservation in a multi-terminal element is to work with the generalizations of Eqs.~(\ref{eq:composite_element_electric_current}) and (\ref{eq:composite_element_spin_current})
\begin{eqnarray}\label{eq:multiterminal_electric_current}
I^e_t &=&  G^e_{tt'} \mu^e_{t'} + C^{\alpha}_{tt'} \tilde\mu^{s\alpha}_{t'},
\\
\label{eq:multiterminal_spin_current}
I^{s\alpha}_t &=&   C^{\alpha}_{t't} \mu^e_{t'} +  S^{\alpha\beta}_{tt'} \tilde\mu^{s\beta}_{t'}.
\end{eqnarray}
The reciprocity requirements translate into $G^e_{tt'} = G^e_{t't}$, $S^{\alpha\beta}_{tt'} = S^{\beta\alpha}_{t't}$, and the indices of $C$ in the second equation being transposed compared with the first.

Conservation of electric current requires $\sum_{t=1}^N I^e_t = 0$, where the summation is performed over all terminals. Two conditions emerge from it
\begin{equation}\label{eq:}
\sum_{t=1}^N G^e_{tt'} = 0,
\qquad
\sum_{t=1}^N C^{\alpha}_{tt'} = 0.
\end{equation}
The first of them is the standard requirement satisfied in any multi-terminal electric element.

\section{No geometric spin ratchets}
While spin electronics may promise various  advantages, spin dissipation hinders spin transmission and is clearly an obstacle. This naturally raises the issue of finding systems with longer spin propagation lengths.

\begin{figure}[t]
  \centering
  \includegraphics[width = 0.3\textwidth]{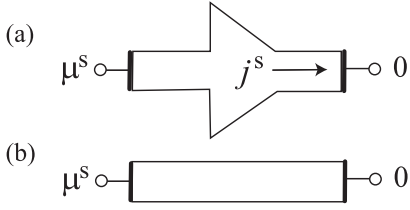}
  \caption{Normal metal elements (a) with directional arrow shape and (b) reference strip.}
  \label{fig:element_Normal_arrow}
\end{figure}

One interesting proposal \cite{abdullah:2014, abdullah:2015} involves optimizing the geometric shape of a conductor -- and a claim that, in an arrow-shaped normal wire (Fig.~\ref{fig:element_Normal_arrow}(a)), spin transmission is enhanced as compared with a rectangular wire (Fig.~\ref{fig:element_Normal_arrow}(b)). Indeed, depending on the precise shape of the arrow, the spin conductance $G^s_{12}$ of the wire with an arrow may or may not be enhanced compared with a rectangular strip. But this does not yet mean that the reason for the enhancement is the orientation of the arrow. The presence or absence of propagation boost due to the geometric asymmetry of the wire should be inferred from a comparison between spin propagation along the arrow direction and opposite to it. And this is precisely where the reciprocity relation (\ref{eq:normal_element_main_result}) applies. It tells us that spin propagation through the arrow-shaped element is the same in both directions. The conductances $G^s_{11}$ and $G^s_{22}$ may differ, and thus the current drawn from the injector may  depend on the side where $\mu^s$ is applied. But, at a given $\mu^s$, the transmitted spin current remains the same regardless of the arrow orientation. We must conclude that an arrow pointing against the spin current flow ``amplifies'' it as much as the one pointing along the flow. This conclusion holds for any passive spintronic element of a kind described above, to which the reciprocity relations apply.

\section{Comparison with the circuit theory formalism}
\label{Sec:comparison_with_circuit_theory}

The ``circuit theory'' (CT) of the Ref.~\onlinecite{nazarov:2000} is a finite-element (lumped element) theory, operating with two types of elementary units:  Normal or ferromagnetic ``nodes'', each characterized by a spatially uniform electron distribution function, and ``contacts'' that define the conductance between the nodes. Spin relaxation may take place  in the nodes, but not in the contacts. A special type of nodes, the ``reservoirs'', set the voltages and spin potentials, applied to the device.

Here we illustrate the correspondence between the matrix conductance $G^{ab}_{tt'}$ of the diffusion-equation description of the preceding sections and the CT matrix conductance. We focus on a simple two-terminal $F/N$ element in the Fig.~\ref{fig:element_Non-equal_C1_C2}. To begin with, the terminology of the two approaches is different: In the diffusion-equation approach, an ``element'' connects two ``contacts'', each characterized by its electric potential $\mu^e$ and spin potential $\mu^{s\alpha}$. In the CT approach, a ``contact'' connects two ``nodes'', each characterized by its $\mu^e$ and $\mu^{s\alpha}$. Thus an ``element'' of the diffusion-equation description should be compared with a CT ``contact'', while a diffusion-equation ``contact'' corresponds to a CT ``node''.

In a CT $F/N$ contact, the electric and spin currents are determined by spin-resolved real conductances $G_{\uparrow}$, $G_{\downarrow}$ and a complex mixing conductance $G_{\uparrow\downarrow}$, which in total makes four real parameters.\cite{nazarov:2000} At the same time, the conductance matrix $G^{ab}_{tt'}$ in the Eq. (\ref{eq:8x8_explicit_matrix}) involves 28 independent parameters. The relation between the parameter sets of the CT and the diffusion-equation description is discussed below.

The settings studied in the Ref.~\onlinecite{nazarov:2000} and in our work are generally different, and comparison is meaningful only where the validity domains of the two approaches overlap. First, the Ref.~\onlinecite{nazarov:2000} assumed no spin dissipation in the contact. Second, it considered an $F$-terminal with the spin potential set to zero, $\mu^{s\alpha}_F = 0$. Without loss of generality, we choose $\mu^e_F = 0$, since currents depend only on the difference $\mu^e_N - \mu^e_F$. Spin potential $\mu^{s\alpha}_N$ of the $N$-contact is allowed to have an arbitrary direction, not necessarily collinear with the magnetization direction $m^{\alpha}$ of the $F$ electrode. Third, the Ref.~\onlinecite{nazarov:2000} studied the currents in the $N$-terminal. Therefore, we shall compare the matrix conductance of a CT contact with that of a diffusive $F/N$ element with spin relaxation lengths $\lambda_{F,N} \to \infty$.

The CT operates with a $2\times 2$ matrix current $\hat I$, related to the electric and spin currents as per $\hat I = (I^e \hat E + I^{s\alpha} \hat\sigma_{\alpha})/2$, where $\hat\sigma_{\alpha}$ are the Pauli matrices. Likewise, the matrix potential is given by $\hat \mu = \mu^e \hat E + \tilde\mu^{s\alpha} \hat\sigma_{\alpha}$. If $z$-axis chosen along $m^{\alpha}$, CT provides the following formula for the current in the $N$-contact
\begin{eqnarray*}
\hat I_N &=& -
\left(\begin{array}{cc}
G_{\uparrow} \mu_{\uparrow\uparrow}(N) & G_{\uparrow\downarrow} \mu_{\uparrow\downarrow}(N)
\\
G^{*}_{\uparrow\downarrow} \mu_{\downarrow\uparrow}(N) &  G_{\downarrow} \mu_{\downarrow\downarrow}(N)
\end{array}\right) =
\\
&=& - \left(\begin{array}{cc}
G_{\uparrow} (\mu^e_N + \tilde\mu^{sz}_N) & G_{\uparrow\downarrow} (\tilde\mu^{sx}_N - i\tilde\mu^{sy}_N
\\
G^{*}_{\uparrow\downarrow} (\tilde\mu^{sx}_N + i \tilde\mu^{sy}_N) &  G_{\downarrow} (\mu^e_N - \tilde\mu^{sz}_N)
\end{array}\right).
\end{eqnarray*}
Recasting this formula in the form $I^a_N = \mathcal{G}^{ab} \tilde\mu^a_N$ one gets the conductance
\begin{equation}\label{eq:CT_4x4}
\hat {\mathcal G} = -
\left|\begin{array}{cccc}
G_{\uparrow} + G_{\downarrow} & 0 & 0 & G_{\uparrow} - G_{\downarrow} \\
0 & 2 Re[G_{\uparrow\downarrow}] & 2 Im[G_{\uparrow\downarrow}] & 0 \\
0 & -2 Im[G_{\uparrow\downarrow}] & 2 Re[G_{\uparrow\downarrow}] & 0 \\
G_{\uparrow} - G_{\downarrow} & 0 & 0 & G_{\uparrow} + G_{\downarrow}
\end{array}\right|.
\end{equation}
The matrix ${\mathcal G}^{ab}$ should be compared with the $4\times 4$ sector $G^{ab}_{NN}$ of the $8 \times 8$ matrix $G^{ab}_{tt'}$ (\ref{eq:G4x4blocks})
\begin{equation}
\label{eq:G_NN_4x4}
G^{ab}_{NN} = \left|\begin{array}{cccc}
-G & C^x_N & C^y_N & C^z_N \\
C^x_N & S^x_N & S^{xy}_N & S^{xz}_N \\
C^y_N & S^{xy}_N & S^y_N & S^{yz}_N \\
C^z_N & S^{xz}_N & S^{yz}_N & S^z_N
\end{array}\right|.
\end{equation}
The sectors $G^{ab}_{NF}$ and $G^{ab}_{FF}$ are not related to ${\mathcal G}^{ab}$.

The $G^{ab}_{NN}$ of the Eq. (\ref{eq:G_NN_4x4}) assumes the form of ${\mathcal G}^{ab}$ in the Eq. (\ref{eq:CT_4x4}) if its entries satisfy a number of conditions:

First, zero entries of $\hat {\mathcal G}$ should be matched by  $C^x_N = C^y_N = S^{xz}_N = S^{yz}_N = 0$. Appendix~\ref{Appendix:symmetry_of_zero_spin_dissipation} shows that this is the case for collinear devices, of which ours is a particular example. Appendix~\ref{Appendix:Conductance of a diffusive F/N contact} illustrates this for the device in the Fig.~\ref{fig:element_Non-equal_C1_C2} by direct calculation along the lines of the Refs.~\onlinecite{nazarov:2000, brataas:2001, brataas:2006, kovalev2002}.

Second, the equality of the first and the last diagonal entries of $\hat {\mathcal G}$ requires $S^{z}_{N} = -G$. Appendix~\ref{Appendix:symmetry_of_zero_spin_dissipation} shows that this property relies both on the collinear character of the device and on the absence of spin relaxation. Under these conditions, the equality $S^z_N = -G$ is protected by a peculiar symmetry of the diffusion equations and boundary conditions. Appendix~\ref{Appendix:Conductance of a diffusive F/N contact} illustrates this by direct calculation and shows that, in the presence of spin relaxation, $S^z_N$ and $-G$ are different.

Third, since the entry $Im[G_{\uparrow\downarrow}]$ appears in $\hat {\mathcal G}$ antisymmetrically, while the symmetry of $\hat G_{NN}$ is our main statement, $\hat {\mathcal G}$ and $\hat G_{NN}$ can be equal only if $Im[G_{\uparrow\downarrow}] = 0$. In terms of the Eq. (\ref{eq:G_NN_4x4}), this means $S^{xy}_N = 0$. In the diffusion-equation description, this can be traced back to the $F/N$ interface being spin-inactive.

To conclude, in the absence of spin relaxation, a diffusive $F/N$ element with spin-inactive interface can indeed be modeled as a single CT contact with symmetric (reciprocal) $\hat \mathcal G$.

\begin{figure}[t]
\center
\includegraphics[width = 0.48\textwidth]{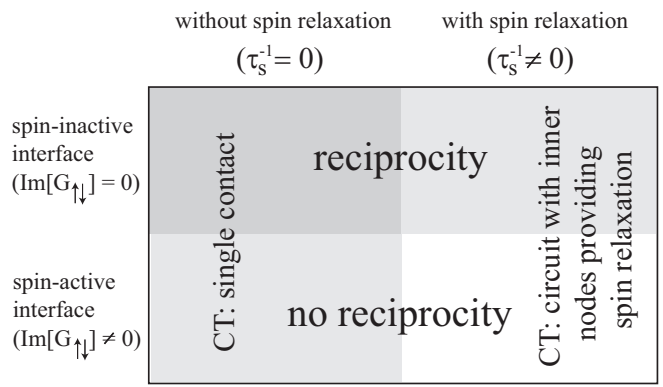}
\caption{Regimes of a diffusive $F/N$ element: Reciprocity properties discussed in our  work break down in the presence of a spin-active interface. In terms of circuit theory, an element can be described as a single CT contact if $\tau_s^{-1} = 0$, but has to be modeled by a CT circuit with spin-relaxing inner nodes when $\tau_s^{-1} \neq 0$.}
 \label{fig:table_of_cases}
\end{figure}

By contrast, in the presence of spin relaxation, $S^{z}_{N} \neq -G$, and thus the $F/N$ element cannot be modeled by a single CT contact of the form (\ref{eq:CT_4x4}). Instead, the model shall involve a CT circuit with at least one inner node, accounting for spin relaxation. All of this is schematically summarized in the Fig.~\ref{fig:table_of_cases}.

Finally, we wish to note that diffusive elements can also be described by equations for the spatially non-uniform $2 \times 2$ spin distribution function.~\cite{hernando:2000} As the node size reduces below the diffusion length, this description explicitly crosses over to the CT formalism in the form of the Ref.~\onlinecite{nazarov:2000}.

\section{Discussion}

In this work, we established reciprocity relations for a class of devices, with or without spin relaxation, where (a) both spin and charge propagate diffusively, (b) the carrier spin aligns itself with magnetization of a ferromagnetic element over a vanishingly short distance, and (c) the $F/N$ interfaces are Ohmic and spin-inactive. Together with charge conservation, the reciprocity relations constrain the form of the conductance matrix: For example, the 64 entries of an $8 \times 8$ conductance matrix of a two-terminal element are reduced to only 28 independent values.

In the case of normal metal elements, reciprocity relations prove impossible ``geometric spin ratchets''\cite{abdullah:2014,abdullah:2015} that would amplify spin current or even transmit it differently in two directions.

Finally, comparison with the circuit theory description of spintronic elements shows that
reciprocity requires spin-active interfaces to be absent.

\acknowledgements{The authors are grateful to Yu.~V.~Egorov, S.~Yu.~Orevkov,
and V.~V.~Schechtman for informative discussions and interest to this work.
Ya.\ B. is grateful to CNRS for financial support, and to the Laboratoire de
Physique Th\'eorique, Toulouse, for the hospitality.}

%=============================================================================
\appendix

\section{Identities} \label{appendix:identities}
For two functions $u({\bf r})$ and $v({\bf r})$, the Gauss theorem gives
\begin{equation}
\int_D (u \Delta v - v \Delta u) dV = \oint_S (u \nabla v - v \nabla u) d{\bf A}
\end{equation}
with $D$ being the integration volume with surface $S$.
Furthermore, for an arbitrary $a({\bf r})$
\begin{equation}\label{eq:identity_scalar}
\int [u \nabla (a \nabla v) - v \nabla(a \nabla u)] dV = \oint [u a \nabla v - v a \nabla u] d{\bf A}.
\end{equation}
This may be generalized for tensors. For $u_{\alpha}({\bf r})$, $v_{\beta}({\bf r})$, and symmetric $A_{\alpha\beta}({\bf r}) = A_{\beta\alpha}({\bf r})$ one has
\begin{eqnarray}
\nonumber
& &
\int [u_{\alpha} \nabla (A_{\alpha\beta} \nabla v_{\beta}) - v_{\alpha} \nabla (A_{\alpha\beta} \nabla u_{\beta})] dV
=
\\ \label{eq:identity_tensor}
& & =
\oint [u_{\alpha} A_{\alpha\beta} \nabla v_{\beta} - v_{\alpha} A_{\alpha\beta} \nabla u_{\beta}]  d{\bf A},
\end{eqnarray}
with repeated index summation assumed.

\section{Conductivity of strong ferromagnets}\label{Appendix:Conductivty_of_strong_ferromagnets}
In a strong ferromagnet with constant magnetization direction $m^{\alpha}$ the vector spin potential satisfies $\mu^{\alpha}(r) = m^{\alpha} \mu^s(r)$, and the spin current satisfies $j^{\alpha}_i(r) = m^{\alpha} j^s_i(r)$. Currents $j^e_i, j^s_i$ and potentials $\mu^e, \mu^s$ are related by the equations
\begin{eqnarray*}
j^e_i &=& - \sigma(r) \nabla_i \mu^e - p(r)\sigma(r) \nabla_i (\mu^s/2),
\\
j^s_i &=& - p(r)\sigma(r)\nabla \mu^e - \sigma(r) \nabla_i (\mu^s/2) .
\end{eqnarray*}
Therefore
\begin{eqnarray}
\nonumber
j^e_i &=& - \sigma(r) \nabla_i \mu^e - p(r)\sigma(r)m^{\alpha} \nabla_i (\mu^{s\alpha}/2),
\\
\label{eq:appendixFM_currents_isotropic}
j^{s\alpha}_i &=& - p(r)\sigma(r) m^{\alpha}\nabla \mu^e - \sigma(r) \nabla_i (\mu^{s\alpha}/2).
\end{eqnarray}
These equations can be combined into
\begin{equation}\label{eq:appendixFM_combined_conductance_notation}
j^a_i = - \Sigma^{ab} \nabla_i \tilde\mu^b
\end{equation}
with $a = \{e,x,y,z\} = \{0,1,2,3\}$, rescaled potentials
$$
\tilde\mu^b = \{\mu^e, \frac{\mu^{sx}}{2}, \frac{\mu^{sy}}{2},  \frac{\mu^{sz}}{2} \},
$$
and a $4 \times 4$ matrix of generalized conductivity
\begin{equation}\label{eq:appendixFM_Sigma_matrix}
\Sigma^{ab} = \left( \begin{array}{cccc}
\sigma & p \, \sigma m^x  & p \, \sigma m^y  & p \, \sigma m^z
\\
p \, \sigma m^x  & \sigma   & 0  & 0
\\
p \, \sigma m^y & 0 & \sigma & 0
\\
p \, \sigma m^z & 0 & 0 & \sigma
\end{array}\right).
\end{equation}
When defined in terms of $\tilde\mu^a$, the generalized conductivity tensor is symmetric.

\section{Symmetry constraints on the conductance matrix}
\label{Appendix:symmetry_of_zero_spin_dissipation}

All of the constrains obtained in this Appendix rely on the device being magnetically collinear. That is, magnetization in all the ferromagnetic parts points along or opposite one and the same direction, denoted as  $m^\alpha$.

\subsection*{In a collinear device, $C^x_N = C^y_N = S^{xz}_N = S^{yz}_N = 0$}

In such a device, all the equations for the electric current density $j^e_i$ and for the component $j^{s z}_i = m^\alpha j^{s \alpha}_i$ of the spin current are invariant with respect to uniform spin rotation $( \mu^{sx} , \mu^{sy}, \mu^{sz} ) \Rightarrow ( \mu^{sx} \cos\varphi - \mu^{sy} \sin\varphi  , \mu^{sy} \cos\varphi + \mu^{sx} \sin\varphi, \mu^{sz} )$ by an arbitrary angle $\varphi$ around the $m^\alpha$. This includes the continuity equations (\ref{eq:electric_current_conservation}, \ref{eq:normal_metal_spin_relaxation}), the expressions (\ref{eq:strong_ferromagnet_electric_current},\ref{eq:strong_ferromagnet_spin_current}) for the currents and the boundary conditions (\ref{eq:electric_boundary_condition},\ref{eq:spin_boundary_condition},\ref{eq:surface_absorption}) at the $F/N$ interface. Therefore, both the electric current $I^e_N$ and the $I^{s z}_N$ component of the spin current are invariant under such a rotation, accompanied by the corresponding rotation $( \mu^{sx}_N , \mu^{sy}_N, \mu^{sz}_N ) \Rightarrow ( \mu^{sx}_N \cos\varphi - \mu^{sy}_N \sin\varphi  , \mu^{sy}_N \cos\varphi + \mu^{sx}_N \sin\varphi, \mu^{sz}_N )$  of the boundary conditions. For the electric current $I^e_N$, the Eq.~(\ref{eq:G_NN_4x4}) yields
$$
I^e_N = - G \mu^e_N + C^x_N \mu^{sx}_N + C^y_N \mu^{sy}_N + C^z_N \mu^{sz}_N .
$$
This expression is invariant with respect to the rotation above only if $C^x_N = C^y_N = 0$.

The same argument for the component $I^{s z}_N$ of the spin current
$$
I^{s z}_N = C^z_N \mu^e_N + S^{x z}_N \mu^{sx}_N + S^{y z}_N \mu^{sy}_N + S^z_N \mu^{sz}_N
$$
leads us to conclude that $S^{x z}_N = S^{y z}_N = 0$.

In a device with non-collinear ferromagnetic parts, the $C^x_N$, $C^y_N$, $S^{x z}_N$ and $S^{y z}_N$ are generally non-zero.

\subsection*{Collinearity and no spin relaxation lead to $G = -S^z_N$}

In a collinear device with no spin relaxation, an extra symmetry guarantees that $G = -S^z_N$.
To see this, we apply potentials $\mu^e_N$, $\mu^{sz}_N$ to the $N$-terminal, while the remaining components of spin potential are set to zero ($\mu^{sx}_N = \mu^{sy}_N = 0$). Using the notation $\tilde\mu^s \equiv \mu^s / 2$ (cf. Eq. (\ref{eq:notations})), the Eqs. (\ref{eq:strong_ferromagnet_electric_current}) and (\ref{eq:strong_ferromagnet_spin_current}) yield
\begin{eqnarray}
\label{eq:strong_ferromagnet_electric_symmetric}
j^e_i &=& - \frac{\sigma}{e^2} \left(  \nabla_i \mu^e
+ p \nabla_i \tilde\mu^{sz}  \right) ,
\\
\label{eq:strong_ferromagnet_spin_symmetric}
j^{sz}_i &=& -\frac{\sigma}{e^2} \left(  \nabla_i \tilde\mu^{sz}
+  p \nabla_i \mu^e \right) ,
\end{eqnarray}
In the normal part, the Eqs. (\ref{eq:normal_metal_electric_current}-\ref{eq:normal_metal_spin_current}) yield
\begin{eqnarray}
\label{eq:normal_metal_electric_symmetric}
j^e_i &=& - \frac{\sigma}{e^2} \nabla_i \mu^e ,
\\
\label{eq:normal_metal_spin_symmetric}
j^{sz}_i &=& -  \frac{\sigma}{e^2} \nabla_i \tilde\mu^{sz} .
\end{eqnarray}

We observe that the system of Eqs. (\ref{eq:strong_ferromagnet_electric_symmetric}-\ref{eq:normal_metal_spin_symmetric}) is invariant under the transmutation $\tilde\mu^e \leftrightarrow \tilde\mu^{s z}$, $j^e_i \leftrightarrow j^{s z}_i$.
Spin relaxation breaks this invariance, since the continuity equations (\ref{eq:electric_current_conservation}) and (\ref{eq:normal_metal_spin_relaxation}) for spin and charge are different. However, as $\tau_s \rightarrow \infty$ in the Eq. (\ref{eq:normal_metal_spin_relaxation}), the symmetry is restored, and the full problem becomes symmetric under the replacement $\tilde\mu^e \leftrightarrow \tilde\mu^{s z}$, $j^e_i \leftrightarrow j^{s z}_i$, completed by interchanging the driving potentials $\mu^e_N$ and $\mu^{sz}_N$ at the N-terminal. All the equations describing the element, including the boundary conditions (\ref{eq:electric_boundary_condition}-\ref{eq:spin_boundary_condition}) at the $F/N$ interface, are invariant under this transformation.

For the total currents $I^e_N$ and $I^{sz}_N$, the Eq. (\ref{eq:G_NN_4x4}) implies
\begin{eqnarray*}
I^e_N &=& - G \mu^e_N + C_N^z \mu^{sz}_N ,
\\
I^{sz}_N &=& C^z_N \mu^e_N + S^z_N \mu^{sz}_N .
\end{eqnarray*}
Upon the symmetry transformation above ($\mu^e_N \leftrightarrow \mu^{s z}_N$, $I^e_N \leftrightarrow I^s_N$), these two equations turn into
\begin{eqnarray*}
I^e_N &=& S^z_N \mu^e_N + C^z_N \mu^{sz}_N ,
\\
I^{sz}_N &=& C_N^z \mu^e_N - G \mu^{sz}_N \ .
\end{eqnarray*}
At the same time, the two equations must remain intact for any $\mu^e_N$ and $\mu^{sz}_N$. This is the case only if $S^{z}_{N} = -G$.

\section{Conductance of a diffusive $F/N$ element}
\label{Appendix:Conductance of a diffusive F/N contact}

\subsection*{Equations and boundary conditions}
Here we consider an $F/N$ element, comprising a thin ferromagnetic wire of length $L_F$ in series with a normal wire of length $L_N$, as shown in the Fig.~\ref{fig:element_Non-equal_C1_C2}. All quantities thus depend only on the coordinate $x$ along the wire, with the origin at the $F/N$ boundary. In the absence of spin-orbit coupling, the spin and coordinate spaces are decoupled. Thus, without loss of generality, we will assume that the magnetization of the F-wire points along the $z$ axis.

Electric and spin potentials are assumed to take zero values at the ferromagnetic terminal
\begin{eqnarray*}
  \mu^e(-L_F) &=& 0 \ , \\
  \mu^{s\alpha}(-L_F) &=& 0 \ .
\end{eqnarray*}
At the normal terminal
\begin{eqnarray*}
  \mu^e(L_N) &=& \mu^e_{N} \ , \\
  \mu^{s\alpha}(L_N) &=& \mu^{s\alpha}_N  .
\end{eqnarray*}
Our goal is to find the currents $j^e_x(L_N) \equiv j^e_N$ and $j^{s\alpha}_x(L_N) \equiv j^{s\alpha}_N$ at the normal terminal. To do this, one has to find potentials $\mu^e(x)$ and $\mu^{s\alpha}(x)$ on the interval $[-L_F, L_N]$. The spin potential obeys equations \cite{rashba_prbrc2000, rashba_epjb2002}
\begin{equation}\label{AppendixC:equation_on_mu_s}
\lambda_{N,F}^2 \frac{d^2{\mu^{s\alpha}}(x)}{dx^2} = \mu^{s\alpha}(x)
\end{equation}
with $\lambda_{N,F}$ being the spin-diffusion lengths in the N- and F-wires. General solutions of these equations can be written in the form
\begin{eqnarray*}
\mu^{s\alpha}(x) &=& X^{\alpha}_{F} e^{x/\lambda_{F}} + Y^{\alpha}_{F} e^{-x/\lambda_{F}}
\quad (-L_F < x \leq 0),
\\
\mu^{s\alpha}(x) &=& X^{\alpha}_{N} e^{x/\lambda_{N}} + Y^{\alpha}_{N} e^{-x/\lambda_{N}}
\quad (0 \leq x < L_N)
\end{eqnarray*}
with coefficients to be determined from the continuity of $\mu^e$, $\mu^s$, $j^e$ and $j^{sz}$ at $x=0$,\cite{brataas:2001, brataas:2006} and from the values of potentials at the terminals.

\subsection*{General solutions}
Since the F-wire is assumed to be a strong ferromagnet, in addition to (\ref{AppendixC:equation_on_mu_s}) spin potential satisfies $\mu^{s\alpha}(x) = m^{\alpha}\mu^s(x)$ in the ferromagnetic part of the element, with $m^{\alpha}$ being the unit vector along the magnetization. Thus we seek spin potential in the F-wire in the form
\begin{eqnarray*}
  \mu^{sx}(x) &=& 0, \\
  \mu^{sy}(x) &=& 0, \qquad \qquad \qquad \qquad (-L_F < x \leq 0)\\
  \mu^{sz}(x) &=& A \sinh \frac{x+L_F}{\lambda_F}.
\end{eqnarray*}
The last expression is written so that it automatically satisfies the boundary condition at $x = -L_F$. Spin potential in the N-wire can be sought in the form
\begin{eqnarray*}
  \mu^{sx}(x) &=& a_x \sinh \frac{x}{\lambda_N}, \\
  \mu^{sy}(x) &=& a_y \sinh \frac{x}{\lambda_N}, \qquad \qquad \qquad (0 \leq x < L_N)\\
  \mu^{sz}(x) &=& a_z \sinh \frac{x}{\lambda_N} + b_z \cosh \frac{x}{\lambda_N}.
\end{eqnarray*}
with unknown $a_{x,y,z}$ and $b_z$. The first two equations ensure the continuity of $\mu^{sx}$ and $\mu^{sy}$ at $x=0$. Matching the spin potentials at the normal terminal we get
\begin{eqnarray}
\label{appendixC:ax}
&&  a_x \sinh \frac{L_N}{\lambda_N} = \mu^{sx}_N, \\
\label{appendixC:ay}
&&  a_y \sinh \frac{L_N}{\lambda_N} = \mu^{sy}_N, \\
\label{AppenidxC:az_bz}
&&  a_z \sinh \frac{L_N}{\lambda_N} + b_z \cosh \frac{L_N}{\lambda_N} = \mu^{sz}_N \ .
\end{eqnarray}
From the continuity of spin potential on the $F/N$ boundary
\begin{eqnarray}
\label{AppendixC:bz_A}
&&  A \sinh \frac{L_F}{\lambda_F} = b_z \ .
\end{eqnarray}
To shorten the expressions in the remainder of this section, we introduce notation
\begin{eqnarray*}
 sh &=& \sinh(L/\lambda), \\
 ch &=& \cosh(L/\lambda), \\
 th &=& \tanh(L/\lambda).
\end{eqnarray*}
The coefficients $a_x$ and $a_y$ are then expressed as
$$
a_x = \frac{\mu^{sx}_N}{sh_N}, \qquad a_y = \frac{\mu^{sy}_N}{sh_N}.
$$
To find $A$, $a_z$, and $b_z$ the conditions of continuity for $\mu^e$, $j^e$ and $j^{sz}$ have to be invoked.

\subsection*{Electric current continuity}
Electric potential in the N-wire obeys the equation
$$
\frac{d^2 \mu^e(x)}{dx^2} = 0 \qquad \qquad (0 \leq x < L_N) \ .
$$
Its solutions are linear functions, so
$$
\mu^e(x) = \mu^e_N \frac{x}{L_N} + \mu^e(0)\left(1 - \frac{x}{L_N} \right)
\qquad (0 \leq x < L_N)
$$
with yet unknown $\mu^e(0)$.

Electric potential equation in the F-wire is more complicated and couples electric and spin potentials. However, its use can be avoided because in the present 1D case the conservation of electric current means $j^e_x = {\rm const} = j^e_N$. Equation (\ref{eq:strong_ferromagnet_electric_current}) then gives a relation for potentials in the F-wire
$$
-\frac{e^2 j^e_N}{\sigma_F} = \frac{d\mu^e}{dx} + \frac{p}{2} \frac{d\mu^{sz}}{dx}
\qquad (-L_F < x \leq 0).
$$
Integrating it from $x = -L_F$ to $x=0$, and using $\mu^e(-L_F) = 0$, gives
\begin{equation}\label{AppendixC:mue_first_expression}
\mu^e(0) = -  \frac{p}{2} \mu^s(0) -\frac{e^2 L_F j^e_N}{\sigma_F}
= -  \frac{p \, b_z}{2} - \frac{e^2 L_F j^e_N}{\sigma_F} \ .
\end{equation}

Electric current flowing through the element can be alternatively expressed by applying formula~(\ref{eq:normal_metal_electric_current}) to the N-wire
\begin{equation}\label{AppendixC:mue_second_expression}
j^e_N = -\frac{\sigma_N}{e^2} \ \frac{\mu^e_N - \mu^e(0)}{L_N}.
\end{equation}
Following the Ref.~\onlinecite{kovalev2002}, we introduce notation
$$
\frac{1}{R} = \frac{\sigma}{e^2 L}  \ .
$$
Combining the Eqs.~(\ref{AppendixC:mue_first_expression}) and (\ref{AppendixC:mue_second_expression}), we find
\begin{eqnarray}
\nonumber
  \mu^e(0) &=& \frac{R_F}{R_N+R_F}\mu^e_N -
  \frac{R_N}{R_N+R_F} \frac{p}{2} b_z \ , \\
\label{AppendixC:je}
  j^e_N &=& - \frac{1}{R_N+R_F} \left(\mu^e_N + \frac{p}{2} \, b_z\right).
\end{eqnarray}

\subsection*{Spin current continuity}
Finally, we use the continuity of $j^{sz}$ at the $F/N$ boundary. Combining Eqs.~(\ref{eq:strong_ferromagnet_electric_current}) and (\ref{eq:strong_ferromagnet_spin_current}) we find in the ferromagnet
$$
j^{sz}(x) = p j^e - \frac{\sigma_F (1-p^2)}{2 e^2} \frac{d\mu^{sz}}{dx} \qquad (-L_F < x \leq 0).
$$
In the normal metal, the Eq.~(\ref{eq:normal_metal_spin_current}) gives
$$
j^{sz}(x) = - \frac{\sigma_N}{2 e^2} \frac{d\mu^{sz}}{dx}  \qquad \qquad \qquad (0 \leq x \leq L_N).
$$
Expressing the derivatives of $\mu^{sz}$ if $F$ and $N$-wires at $x=0$ in terms of the unknown coefficients, we get the continuity condition
$$
- \frac{\sigma_N}{e^2\lambda_N} \frac{a_z}{2} \,  = p j^e_N - \frac{\sigma_F}{e^2\lambda_F}  (1-p^2) \frac{ch_F}{2} A \ .
$$
Substituting the electric current from (\ref{AppendixC:je}), we recast the preceding equation in the final form
\begin{equation}\label{AppendixC:az_bz_A}
\frac{L_N}{\lambda_N R_N} \frac{a_z}{2}  = \frac{p (\mu^e_N + p b_z/2)}{R_N+R_F} + \frac{L_F (1-p^2)}{\lambda_F R_F}  \frac{ch_F}{2} A
\end{equation}

\subsection*{Solving for unknown coefficients}
Eqs.~(\ref{AppenidxC:az_bz}), (\ref{AppendixC:bz_A}), and (\ref{AppendixC:az_bz_A}) can be now solved to give the unknown coefficients.
The results can be presented in a more compact way using the notation
\begin{eqnarray*}
&& t = \frac{\lambda}{L} \tanh\frac{L}{\lambda}, \quad
s = \frac{\lambda}{L} \sinh\frac{L}{\lambda} \ ,
\\
&& \frac{1}{R_{\rm eff}} = \frac{1}{R_N t_N} + \frac{p^2}{R_N + R_F} + \frac{1-p^2}{R_F t_F} \ .
\end{eqnarray*}
This gives
\begin{eqnarray}
\nonumber
a_z &=& \frac{2 p R_{\rm eff}}{R_N + R_F} \frac{\mu^e_N}{th_N}
+ \left(1 -  \frac{R_{\rm eff}}{t_N R_N} \right)\frac{\mu^{sz}_N}{sh_N} \ ,
\\
\label{AppendixC:az_bz_formulas}
b_z &=& - \frac{2 p R_{\rm eff}}{R_N + R_F} \mu^e_N +
\frac{R_{\rm eff} }{s_N R_N} \mu^{sz}_N.
\end{eqnarray}

\subsection*{Currents at the normal terminal}
Spin currents at the normal terminal ($x = L_N$) are given by
\begin{eqnarray*}
j^{sx}_N &=& -\frac{L_N}{2 \lambda_N R_N} a_x ch_N,
\\
j^{sy}_N &=& -\frac{L_N}{2 \lambda_N R_N} a_y ch_N,
\\
j^{sz}_N &=& -\frac{L_N}{2 \lambda_N R_N} (a_z ch_N + b_z sh_N).
\end{eqnarray*}
Substituting the expressions for $a_{x,y,z}$ and $b_z$ into the equations above, we get
\begin{eqnarray}
\nonumber
j^{sx}_N &=& -\frac{1}{R_N t_N} \frac{\mu^{sx}_N}{2},
\\
\label{AppendixC:spin_current_answers}
j^{sy}_N &=& -\frac{1}{R_N t_N} \frac{\mu^{sy}_N}{2},
\\
\nonumber
j^{sz}_N &=& - \frac{1}{R_N} \left( \frac{1}{t_N} - \frac{R_{\rm eff}}{s_N^2 R_N} \right)  \frac{\mu^{sz}_N}{2}
\\
\nonumber
&-& \frac{p R_{\rm eff}}{s_N R_N (R_N + R_F)}\mu^e_N \ .
\end{eqnarray}
Electric current is obtained by substituting expression (\ref{AppendixC:az_bz_formulas}) for $b_z$ into Eq.~(\ref{AppendixC:je})
\begin{eqnarray}
\nonumber
j^e_N &=& - \frac{1}{R_N + R_F} \left( 1 - \frac{p^2 R_{\rm eff}}{R_N + R_F} \right) \mu^e_N
\\
\label{AppendixC:electric_current_answer}
&-& \frac{p R_{\rm eff}}{s_N R_N (R_N + R_F) } \frac{\mu^{sz}_N}{2} \ .
\end{eqnarray}
For completeness we also give an expression for
\begin{eqnarray*}
\mu^e(0) &=& \left( \frac{R_F}{R_N + R_F} + \frac{p^2 R_N R_{\rm eff}}{(R_N + R_F)^2} \right) \mu^e_N
\\
&-& \frac{p R_{\rm eff}}{s_N(R_N + R_F)} \frac{\mu^s_N}{2} \ .
\end{eqnarray*}

\begin{figure}[t]
\center
\includegraphics[width = 0.4\textwidth]{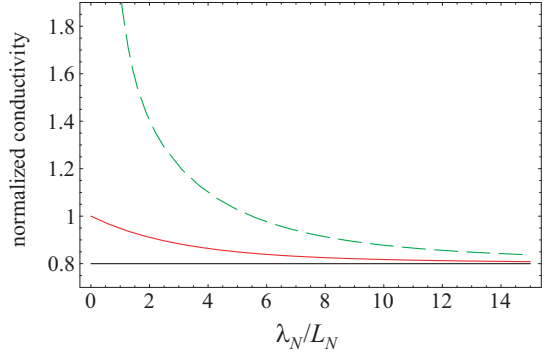}
\caption{(Color online)Conductances $G/G_0$ of the Eq.(\ref{eq:G}) (red/gray), $-S^z_N/G_0$ of the Eq.(\ref{eq:SNz}) (green/gray-dashed), and their limiting value $G_{lim}/G_0$ of the Eq.(\ref{eq:Glim}) (black), normalized to $G_0 \equiv G|_{p=0} = 1/(R_F + R_N)$ and plotted as a function of $\lambda_N/L_N$. For illustrative purposes, other parameters are set to $\lambda_F/L_F = 0.1 (\lambda_N/L_N)$ and  $R_F = 0.1 R_N$.}
 \label{fig:GvsLambda}
\end{figure}

\subsection*{Conductance matrix}
Using (\ref{AppendixC:spin_current_answers}) and (\ref{AppendixC:electric_current_answer}) we can write down the entries of the $4 \times 4$ sector $G^{ab}_{NN}$, Eq.~(\ref{eq:G4x4blocks}). Recall here that they are defined so that currents are positive when they flow out of the element. The nonzero entries are
\begin{eqnarray}
\label{eq:G}
G & = & \frac{1}{R_N + R_F} \left( 1 - \frac{p^2 R_{\rm eff}}{R_N + R_F} \right),
\\
C_N^z &=& -\frac{p R_{\rm eff}}{s_N R_N (R_N + R_F) },
\\
S_N^x &=& S_N^y = -\frac{1}{t_N R_N},
\\
\label{eq:SNz}
S_N^z &=&  - \frac{1}{R_N} \left( \frac{1}{t_N} - \frac{R_{\rm eff}}{s_N^2 R_N} \right) .
\end{eqnarray}
The remaining ones are equal to zero, in accordance with the explanations of Appendix~\ref{Appendix:symmetry_of_zero_spin_dissipation}.

\subsection*{Limit of zero spin dissipation}

The limit of zero spin dissipation corresponds to infinite spin diffusion lengths
$\lambda_{N,F} \to \infty$. In this limit, $s_{N,F} \to 1$, $t_{N,F} \to 1$, and thus
$$
\frac{1}{R_{\rm eff}} \to \frac{1}{R_N} + \frac{p^2}{R_N + R_F} + \frac{1-p^2}{R_F}
$$
or
$$
R_{\rm eff} \to \frac{R_N R_F (R_N + R_F)}{(R_N + R_F)^2 - p^2 R_N^2} \ .
$$
Using these properties we find that conductances $G$ and $-S_N^z$ indeed approach the same limit (see Fig.~\ref{fig:GvsLambda})
\begin{equation}
\label{eq:Glim}
- S_N^z, G \to G_{lim} = \frac{(R_N + R_F) - p^2 R_N}{(R_N + R_F)^2 - p^2 R_N^2} \ .
\end{equation}
This expression for $G_{lim}$ reproduces the results of Refs.~\onlinecite{nazarov:2000} and \onlinecite{brataas:2001}, where spin diffusion length was set to infinity from the outset, and thus completes the correspondence between $G^{ab}_{NN}$ and ${\mathcal G}^{ab}$, as discused in the Sec.~\ref{Sec:comparison_with_circuit_theory}.

\end{document}